# Title: Towards Atom-by-Atom Fabrication: Mechanosynthetic donation and abstraction

**Authors:** Brandon Blue*, Mathieu Morin*, Alex Inayeh, Rosemary Cranston, Cameron J. Mackie, Marc Savoie, Adam Bottomley, Christian J. Imperiale, Zehra Ahmed, Rafik Addou, Aly Asani, Eduardo Barrera-Ramirez, Jeremy Barton, Doreen Cheng, Megan Cowie, Chris Deimert, Tyler Enright, James Zhangming Fan, Robert A. Freitas Jr., Alan T.K. Godfrey, Ryan Groome, Si Yue Guo, Kareem A. Clarcia, Aru Hill, Taleana Huff, Mark Jobes, Robert J. Kirby, Sam Lilak, Hadiya Ma, Adam C. Maahs, Oliver MacLean, Steven M. Maley, Michael Marshall, Terry McCallum, Ralph Merkle, Matthew Moses, Jonathan Myall, Ryan Plumadore, Adam Powell, Henry Rodriguez, Sam Rohe, Luis Sandoval, Khalil Sayed-Akhmad, Benjamin Scheffel, Kashif Tanveer, Bheeshmon Thanabalasingam, Denis A.B. Therien, Janice L. Wong, Reid Wotton, Cristina Yu, Damian G. Allis, Michael Drew, Matthew R. Kennedy, Tait Takatani, Marco Taucer, Dušan Vobornik, Ryan Yamachika, Mathieu Durand

**Affiliations:**

[1]*CBN Nano Technologies, Inc.; Ottawa, K1Y 4W5, Canada*

*Corresponding authors.

**Abstract:** Enabled by inverted-mode scanning tunneling microscopy (IM-STM) and the use of functionalized molecular tools, we demonstrate positionally-controlled mechanosynthetic addition (donation) of carbon and subtraction (abstraction) of silicon atoms on a model build site: atomically clean and crystalline Si(100). The resulting structures represent the first demonstrations of an emerging ability to manipulate radical chemistry with positional control of specific atoms and moieties in 3D. Furthermore, by comparing the behavior of molecular tools designed for atomic donation versus abstraction, we highlight general principles governing molecular tool design for selective and reliable mechanosynthetic functionality.

## Main Text

The state-of-the-art in atomically precise 3D manipulation has generally revolved around scanning probe microscopy (SPM) based manipulation of atoms, especially hydrogen on H:Si(100) surfaces (*1*), often under electronic bias (*2*). Other advances have been made in selective, bias-driven removal of moieties from small, atomically-well defined patches in H:Si(100) (*2*) or 3D molecules (*3*). These techniques rely on metal probes, often with poor control over the atomic arrangement, requiring CO or other functionalization methods (*4–6*) to improve their resolution and reproducibility for specific tasks. Furthermore, the use of bias to mediate these reactions includes influences on the local electric field from the mesoscopic shape of the probe supporting the apex atom or moiety, especially on semiconducting surfaces (*7*). These restrict the scope of feasible structures and suitable precursors for follow-up device processing and limit the effective resolution of patterns.

Recently published work in inverted-mode scanning tunneling microscopy (IM-STM) with molecular tools offers a path towards atomically precise addition (donation) and subtraction (abstraction) of individual atoms from a well-defined build site, which we refer to as atomically precise fabrication (APF) *via* 'positionally-controlled mechanosynthesis' (*8*). Mechanosynthesis, in this sense, denotes the synthesis of specific atomic structures by the mechanical manipulation of precursors in 3D space with precise, positional control over covalent bond formation. With this approach, no bias is needed during the critical steps of the reaction, and all reactions are done *in situ*, without exposure to gaseous precursors. Here, we use this bottom-up process to generate a panoply of previously unreported carbon and silicon structures, which highlight the potential of this method for the atomically-precise fabrication of complex and covalently-assembled products.

## Molecular Tool Design

Adamantane molecular tools are generally described in three parts: 1) flexible 'legs' responsible for surface binding by chemisorption of the 'feet'; 2) an adamantane core to support the desired orientation; and 3) the 'feedstock' at the bridgehead position which will take part in the reaction (*9*, *10*). Synthesis of this class of molecular tools has been described recently (*11*). The **e**thyl-legged, **a**damantane molecular tool with **o**xygen-terminated feet, a **ge**rmanium bridgehead atom, and a **–$C_2$I** terminal functional group as feedstock is referred to by a common name: **EAOGe–$C_2$I** (Fig. 1A/B). When all three feet bind to a surface, the tripodal geometry ensures that the feedstock is available (Fig. 1C). EAOGe–$C_2$I's feedstock is connected to the core *via* a single C–Ge σ-bond (bond dissociation enthalpy, BDE = 5.213 eV, see *Computational Methods Details,* Fig. S1) at the bridgehead position. The iodine atom protects the reactive feedstock until it is deiodinated (*9*, *12*), which generates a long-lived ethynyl radical (–$C_2^{\bullet}$, where $^{\bullet}$ represents any moiety's radical character) under ultra-high vacuum (UHV) conditions (*13*). Variants are synthetically accessible, including other leg lengths, bridgehead atoms, and feedstock moieties (*11*), also shown in the *Improved Selectivity* section below.

## Enabling Atomically Precise Chemistry

Control over both sides of the tunnel junction is crucial to enable positionally-controlled mechanosynthesis. On the probe side, our silicon probe chips (SPCs) present an atomically flat Si(100) mesa, including the >5x5 nm$^2$ build site where the molecular tool interactions occur (*8*). On the sample side, molecular tools are vapor-deposited on a conventionally-prepared Si(100) wafer (*14*), where the optimal molecular orientation presents an accessible functional group. The tall, sharp molecules on the sample surface generate reflected probe images (RPIs) of the comparatively broad silicon apex, yielding an inverted tunneling architecture; in IM-STM, the individual, surface-bound molecular tools act as local probes of the SPC's apex mesa (Fig. 1A). Each tool thus generates its own RPI, and the deposition density must be controlled to minimize overlap of RPIs. Though not all tools bind in this manner (*13*), formation of a well-resolved RPI is a strong prerequisite heuristic which selects for properly-oriented tools; improperly-bound tools are readily avoided.

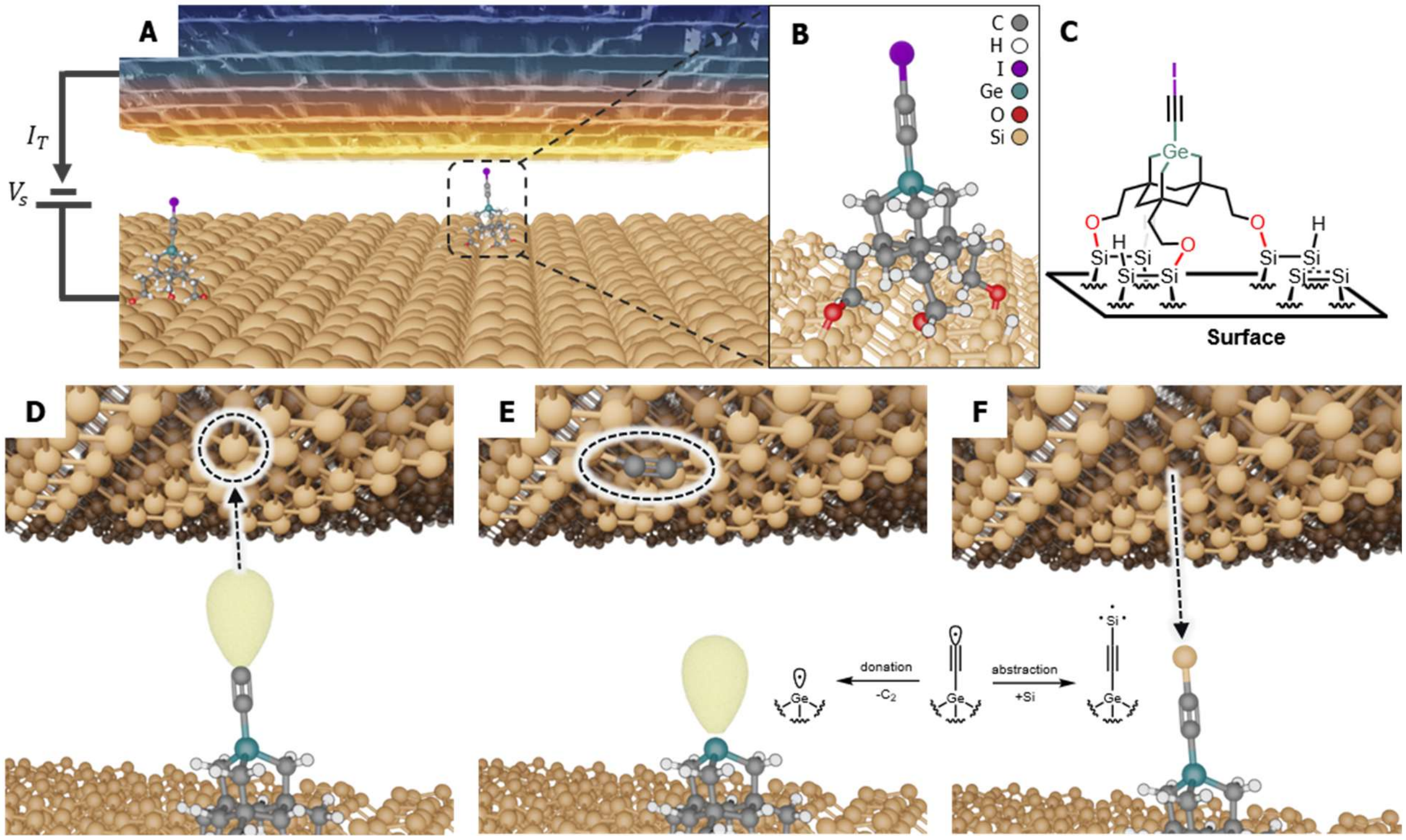

**Fig 1. The molecular tool and general mechanosynthetic processes.** (**A**) Overview of the probe and sample (above and below, respectively, not to scale) for IM-STM enabled mechanosynthesis. (**B**) Zoomed-in focus on the desired surface binding of the EAOGe–$C_2$I molecular tool. (**C**) Chemical scheme for (B). (**D**) The deiodinated tool (EAOGe–$C_2^\bullet$). Approaching the tool to the probe surface and retracting may cause either mechanosynthetic (**E**) donation, yielding EAOGe$^\bullet$ or (**F**) abstraction, yielding EAOGe–$C_2$Si$^\bullet$.

## General Mechanosynthesis Methodology

Here, we outline the typical process of mechanosynthesis under cryogenic (4 K) ultra-high vacuum conditions (see *Materials and Methods* also). First, an appropriate molecular tool on the sample surface is identified by its RPI, followed by selecting a suitable build site on the SPC (Fig. 1A). The molecular tool is then deiodinated *via* electronic bias (*8*), generating an ethynyl radical (Ge–$C_2^\bullet$) in the case of –$C_2$I terminated tools (Fig. 1C/D), which is done away from the build site to avoid contamination of this region by the iodine. Deiodination produces a characteristic, discontinuous increase in the STM baseline tunneling position ($\Delta z$, ~300 pm), and a subsequent change in the IM-STM appearance of the RPI. These changes mirrored those of other functionalized-probe observations (*4*, *15*), and were consistently observed for all successful deiodinations to reliably identify the state of the feedstock.

The EAOGe–$C_2^{\bullet}$ tool is then guided with sub-ångström precision to the desired reaction site by translation of the SPC relative to the tool's RPI. Reaction is initiated by pausing STM *z*-axis feedback and extending the SPC by a set distance towards the tool without applied bias (see *Materials and Methods*). As the SPC and tool approach one another, a new covalent bond is formed between the –$C_2^{\bullet}$ and a surface silicon atom. This Si–C σ-bond (BDE = 5.184 eV) is opposed to the Ge–C σ-bond (BDE = 5.213 eV) at the bridgehead. Upon retraction, the expected outcomes based on these BDEs are either Ge–C homolytic bond cleavage, resulting in a germyl radical-terminated molecular tool (EAOGe$^{\bullet}$) (Fig. 1E) on the silicon sample and a transient pendent ethynyl radical on the probe apex, or breaking of the Si–C σ-bond, reverting the tool to its initial EAOGe–$C_2^{\bullet}$ state. The very small BDE difference between both and the notable limitations of the model theoretical system (see *Computational Methods*) complicate product prediction.

Silicon abstraction is another possible outcome (Fig. 1D/F), where upon retraction the silicon remains covalently bound to the $C_2$ moiety, and is thus abstracted from the build site (Si–Si BDE = 2.18–2.59 eV) (*16*). With three such bonds per surface silicon, the abstraction becomes more favorable if bonds are broken sequentially. The exact chemical state of the silylated tool (Fig. 1F) is uncertain, though it is sufficiently stable to image the SPC (see *Imaging Modality*). If no change was observed, the interaction was repeated with iteratively deeper extensions until a reaction occurred (see *Materials and Methods*). Control and characterization of the interaction outcome are important aspects towards developing mechanosynthetic applications. Other factors influencing the ability to select for a desired outcome (*e.g.* interaction site selection, *etc.*) are described in the *Supporting Information*.

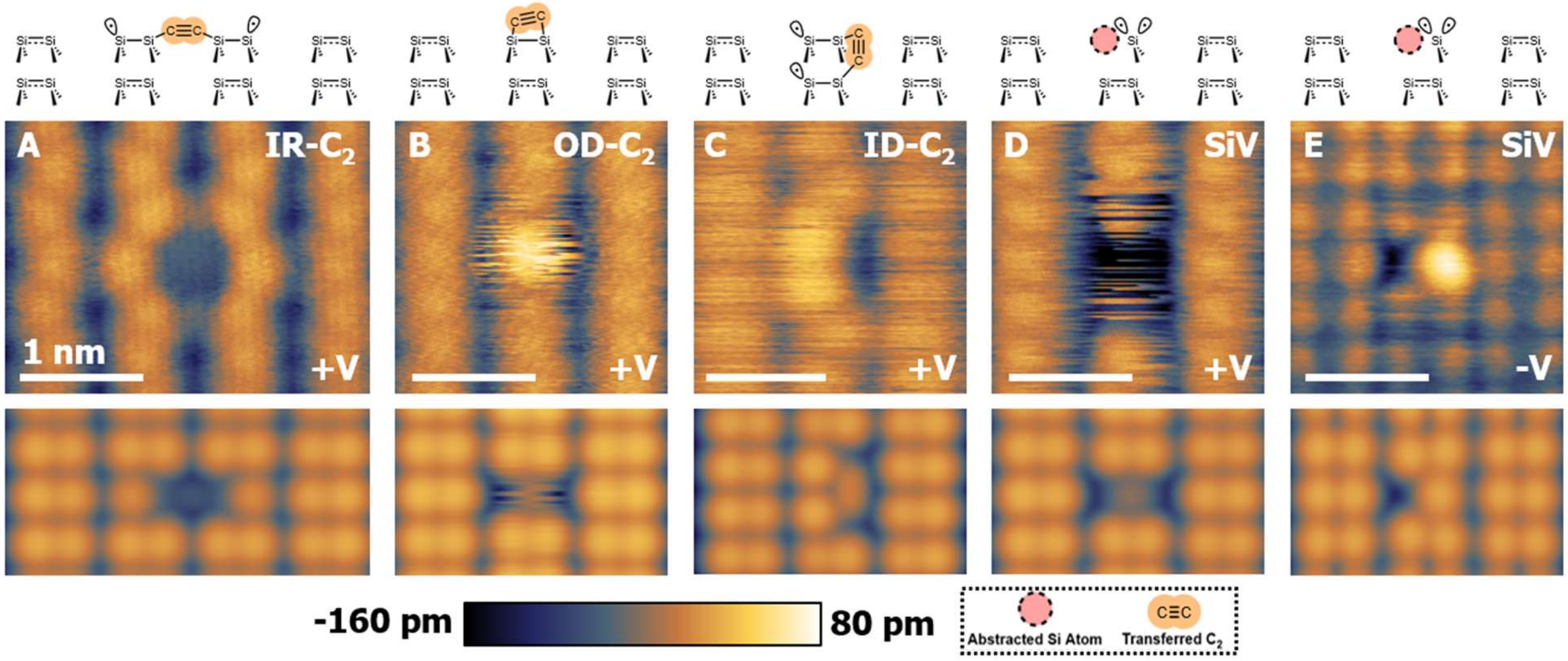

**Fig 2. Individual donation and abstraction mechanosynthetic transfers.** Experimental and simulated images of single interaction outcomes with EAOGe–$C_2$I. (**A**) Inter-row (IR-$C_2$), (**B**) on-dimer (OD-$C_2$), and (**C**) inter-dimer (ID-$C_2$) donation products, compared to (**D**) a single silicon vacancy (SiV). Each outcome is paired with proposed schematic arrangements (top row) with respect to the Si(100) lattice. (**E**) An aggregate image highlighting one of the minima of the SiV feature. Images gathered in IM-STM with EAOGe–$C_2$I tools under (A-D) +3.1 to +3.5 V, and (E) −3.3 V for applied sample bias and 10 to 20 pA current tunneling conditions. STM images simulated at ±2.8 V bias with 50 nA isocurrent value. Simulation for (B) includes post-processing to account for switching, see *Supporting Information* for details.

**Observation Of Atomically Precise Material Transfer to the Si Probe Apex**

Homolytic cleavage of C–Ge during retraction yields a pendent ethynyl radical expected to be transient, spontaneously forming a second Si–C σ-bond with the SPC surface. Subsequent IM-STM imaging with unperturbed EAOGe–$C_2$I yielded one of two major features consistent with carbon dimer donation, though precise bond order cannot be determined from IM-STM images alone: inter-row $C_2$ (IR-$C_2$, Fig. 2A), or on-dimer $C_2$ (OD-$C_2$, Fig. 2B). A third, much rarer inter-dimer $C_2$ (ID-$C_2$) outcome was also observed (Fig. 2C). The same three structures were observed in previous work showing electron-induced dehydrogenation of acetylene ($C_2H_2$) and ethylene ($C_2H_4$) on Si(100) (*17*). Under typical IM-STM imaging conditions, the expected $C_2$ π-bonds appeared relatively dark compared to the surface silicon atoms, while the adjacent silicon dangling bonds appeared similarly bright to the other, unreacted dimers (Fig. 2A). The OD-$C_2$ configuration, with the $C_2$ subunit sitting directly atop but slightly off-center from the dimer (Fig. 2B), appeared relatively brighter than the neighboring silicon sites. OD-$C_2$ features exhibited dynamic switching during IM-STM imaging, seen in Fig. 2B as localized streaking and instability in the immediate vicinity of the feature, perhaps analogous to other instances of dynamic dimer buckling (*17*) or due to forces on the imaging molecule itself, as in the case of CO-functionalized SPM tips (*18*).

In many instances, a silicon vacancy was formed (SiV, Fig. 2D) rather than donating a carbon. Sequential bond breaking depends strongly on the relative positioning and motion of the individual atoms throughout a given interaction: referred to as that interaction's trajectory, (see *Improved Selectivity* and Fig. S2). When surrounded by intact silicon dimers, SiV features exhibited a characteristic dynamic switching with several conformations (Fig. S3), including destabilization of neighboring dimers in the same row. Such instability was not noted on the hydrogenated surface for subsurface vacancies (*19*). One of these minima is demonstrated as Fig. 2E, where the minima state was extracted from individual line scans across multiple, repeated images consistent with static simulations. The assignment of the SiV feature as a silicon atom vacancy was further supported by distinct imaging characteristics of the molecular tool used, (collectively referred to as the tool's imaging modality), and subsequent reconstructions of the silicon build site surface. Opposite bias IM-STM images of each feature presented in Fig. 2 are found in the *Supporting Information* (Fig. S4).

**Imaging Modality as a Characterization Tool**

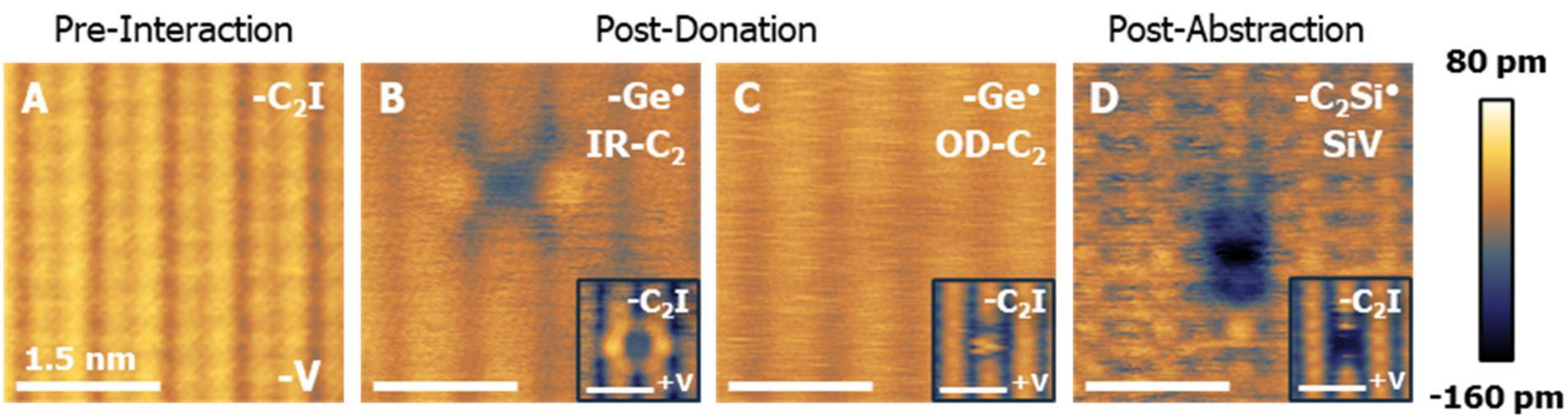

**Fig 3. IM-STM imaging modality relationship to molecular tool state.** Imaging modalities observed using EAOGe–$C_2$I (**A-D**) for (A) the pre-interaction –$C_2$I terminated tool, the post-donation –Ge$^{\bullet}$ terminated tool resulting in (B) an IR–$C_2$ and (C) an OD-$C_2$, and (D) the post-abstraction –$C_2$Si$^{\bullet}$ terminated tool. Insets: Images of the feature from a reference –$C_2$I imager for comparison, with scale bars of 1.5 nm. Images gathered in IM-STM under −3.5 to −3.7 V applied

sample bias for main panel images and +3.2 to +3.5 V for inset images, and 10 to 20 pA current tunneling conditions.

The influence of probe-shape artifacts and tip functionalization on traditional SPM techniques is well known, with certain conditions allowing for the extraction of chemical information from SPM (*4*, *15*). In the case of IM-STM with molecular tools, however, a variety of other factors can confound the analysis of a given image without context. The available biases are restricted by the relative band-gaps of the probe and substrate materials chosen, which vary based on annealing time in the case of arsenic-doped Si(100) (*20*). The effective radius of curvature of the SPC also changes with repeated annealing (*e.g.*, influencing tip-induced band-bending (*7*, *8*, *21*)). The chemical stability of the molecular tools themselves (especially to charge transfer) during tunneling must also be considered.

Thus, a concrete demonstration of the utility of IM-STM imaging modality as a chemical signature was necessary to better characterize more complex mechanosynthetic products by calibrating against simpler base features. Imaging modalities are generally distinguished by relative resolution, presence or absence of contrast inversion, and apparent symmetries in the RPI's Si(100) lattice sites, similar to factors noted by others in the case of qPlus probe shape effects (*22*, *23*). These are primarily dependent on the chemical nature and physical orientation of the terminal moiety of the tool, providing a strong basis for identification of the end-state of the tool in instances where the nature of the product is not immediately self-evident (*e.g.,* when atomic rearrangements obscure the mechanosynthetic outcome).

Fig. 3 highlights typical modalities for EAOGe–$C_2$I (Fig. 3A) and its most-observed variants achieved after mechanosynthetic interaction: EAOGe$^{\bullet}$ for donation (Fig. 3B), and EAOGe–$C_2$Si for abstraction (Fig. 3D). Across hundreds of mechanosynthetic interactions, the resulting post-interaction modalities were collated and categorized by their products *via* imaging with independent EAOGe–$C_2$I tools (Insets Fig. 3B-D), revealing clear correlations. The EAOGe–$C_2^{\bullet}$ modalities were also instructive, but the radical itself was generally less stable, restricting the available biases suitable for IM-STM imaging. Modality trends were found, as expected, to be insensitive to the length of the tools' leg moieties (Fig. S5). Modalities can also be categorized and compared by various image analysis techniques, including Fast-Fourier Transforms (FFT (*24*), see *Supporting Information* Fig. S6) or machine-learning approaches (*25*).

**Multi-Transfer Mechanosynthesis Sequences**

Having demonstrated the fundamental elements of mechanosynthesis, we sought to generate new surface features *via* multiple transfers near one another at the same build site. Simple, *z*-only trajectories under these conditions, though, were generally limited to a selectivity of ~40-60% with a notable interaction site dependence (Fig. S2). These limitations were broadly associated with the unpredictability of the pendent $C_2$'s relaxation (when intending specific donation) and the weak Ge–C bond (when intending abstraction). This limited the ability to target specific arrangements to ~5 or fewer sequential mechanosynthetic interactions. Though individual outcomes are difficult to control, these opportunistically-identified products are instructive.

In the case of SiV, for example, the dynamic nature of the isolated feature is very apparent (Fig. 2C, Video S1, and Fig. S7), even at 4 K. However, a second silicon abstraction event at the same dimer site yields a static silicon dimer vacancy ('DV', Fig. 4A), appearing as a dark depression, consistent with native dimer vacancy defects (*26*). When the second silicon abstraction is instead performed on the adjacent dimer of a neighboring row, the two adjacent SiV features undergo spontaneous rearrangement to form a new reconstructed dimer (RD) spanning the previous trough position, offset by half unit cell (Fig. 4B). Similarly, multiple reconstructed dimers (MRD, Fig. 4C) occurred when the two in-line vacancies were further separated. In the 3-MRD example shown, the dimers appeared to spontaneously reconstruct mid-scan, perhaps driven by the electric field and sustained current present during STM imaging.

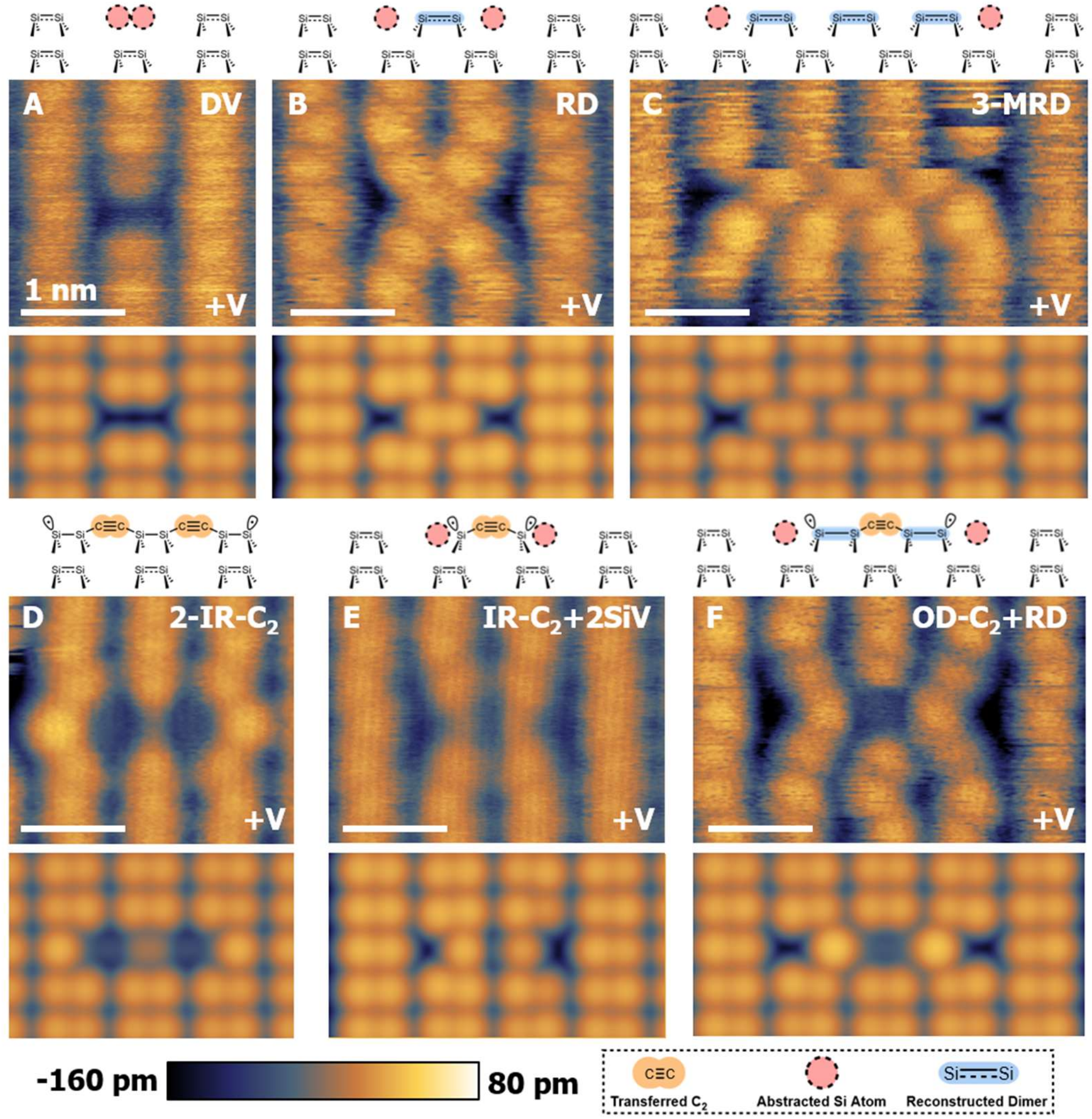

**Fig 4. Multiple donation and abstraction mechanosynthetic transfers.** Experimental and simulated images of multi-interaction mechanosynthetic outcomes with EAOGe–$C_2$I. Multiple abstraction products include (**A**) silicon divacancy (DV), (**B**) reconstructed dimer (RD), and (**C**) multiple reconstructed dimers (3-MRD). Multiple donation products include (**D**) adjacent IR-$C_2$ donations, with combined abstraction and donation products including an IR-$C_2$ with (**E**) two single vacancies to either side, and (**F**) a reconstructed dimer. Each outcome is paired with a proposed schematic configuration with respect to the Si(100) lattice. Images gathered in IM-STM with EAOGe–$C_2$I tools under +3.0 to +3.5 V bias and 10−20 pA current tunneling conditions. STM images simulated at +2.8 V sample bias with a 50 nA isocurrent.

Abstraction and donation events can also, in principle, be combined towards more advanced structures. Adjacent IR-$C_2$ donations across neighboring troughs (2-IR-$C_2$, Fig. 4D) appeared darker in the center, (where the central dimer is fully satisfied), than the simplest, trivial case of summing two, independent IR-$C_2$ features. In another case, the neighboring silicon atoms were abstracted from either side of an initial IR-$C_2$ feature. Each proposed silicon abstraction resulted in the expected –$C_2Si^{\bullet}$ modality immediately after transfer. The final product appears as a trough-centered $C_2$ with silicon vacancies to either side, consistent with simulations (IR-$C_2$+2SiV, Fig. 4E). Similar silicon abstractions were performed at adjacent dimers on the neighboring rows to either side of an OD-$C_2$ feature. The adjacent SiV features caused the splitting of the silicon dimer beneath the $C_2$ subunit and resulted in dimer reconstruction to either side (OD-$C_2$+RD, Fig. 4F). Due to the configuration of the surrounding silicon atoms, the resulting feature bears an apparent similarity to the IR-$C_2$ feature despite being centered on-dimer. A complete overview of the mechanosynthetic steps and intermediates for Fig. 4F are shown in Fig. S8. See Fig. S9 for opposite bias IM-STM images of each feature presented in Fig. 4.

**Improved Selectivity for Si Atom Abstraction *via* Molecular Design**

Further exploring the limits of mechanosynthetic selectivity, under these *z*-only conditions with the silicon dimer center as the interaction point, donation occurred at a rate of 107/197 interactions with EAOGe–$C_2$I. This statistical sampling corresponds to an estimated true donation rate of ~54±7% with 95% confidence following Wilson's method (*27*). In addition to selecting for additive fabrication (as in our companion paper to this work (*28*)), we anticipated that selecting for abstraction would also be desirable for subtractive silicon patterning. We thus developed a specific molecular tool to favor abstraction and minimize donation as an undesirable side reaction for that application.

By replacing the germanium bridgehead atom with carbon in the molecular tool synthesis step, we hypothesized that the C–C bond (BDE = 5.488 eV) would further favor silicon atom abstraction. Additionally, ethyl-alcohol leg groups were replaced with methyl-alcohol leg groups to minimize the number of accessible surface bound configurations (*11*). This new molecular tool, MAOC–$C_2$I, demonstrated a remarkable ability to abstract silicon atoms, with no observed $C_2$ donations. Instead, abstraction was the only outcome when following *z*-only, on-dimer trajectories with MAOC–$C_2$I, both at 4 K (63/63 interactions) and 77 K (100/100 interactions), further supporting the interpretation of the SiV feature as silicon abstraction. See Figures S11-S28 for a detailed synthetic scheme and full characterization of MAOC–$C_2$I .

The improvement to perfect selectivity for abstraction with the simple exchange of a single atom in the tool's structure is a strong indicator of the importance of molecular tool design for specific applications. Furthermore, this perfect selectivity allowed us to demonstrate more extensive subtractive patterning of silicon, even at 77 K, demonstrating the robustness of the method. DV pairs, for example, were readily generated both cross-trough (Fig. 5A) and along-row (Fig. 5B), identical to native defects (*26*). In some instances, a single interaction could generate a DV with the second silicon remaining on the SPC surface as a mobile silicon adatom (Fig. 5C), distinct from ad-dimers simulated by others (*29*). Such adatoms could then be mechanosynthetically abstracted by the same *z*-only interaction process, where desired, as a preliminary form of error correction (Fig. 5C). These capabilities were combined to generate structures of ~10 sequentially abstracted silicon atoms shown in Fig. 5D, (see Fig. S24 for opposite bias imaging). Comparisons to simulations (Fig. 5E) suggest the opening of a new, subsurface trough orthogonal to the surface in the newly uncovered region, akin to Si(100) step edges. Surface dynamics in the presence of such precisely engineered defects are expected to play an important role in future APF efforts, and may lead to future distinctions between mechanosynthetic outcomes at varying temperatures.

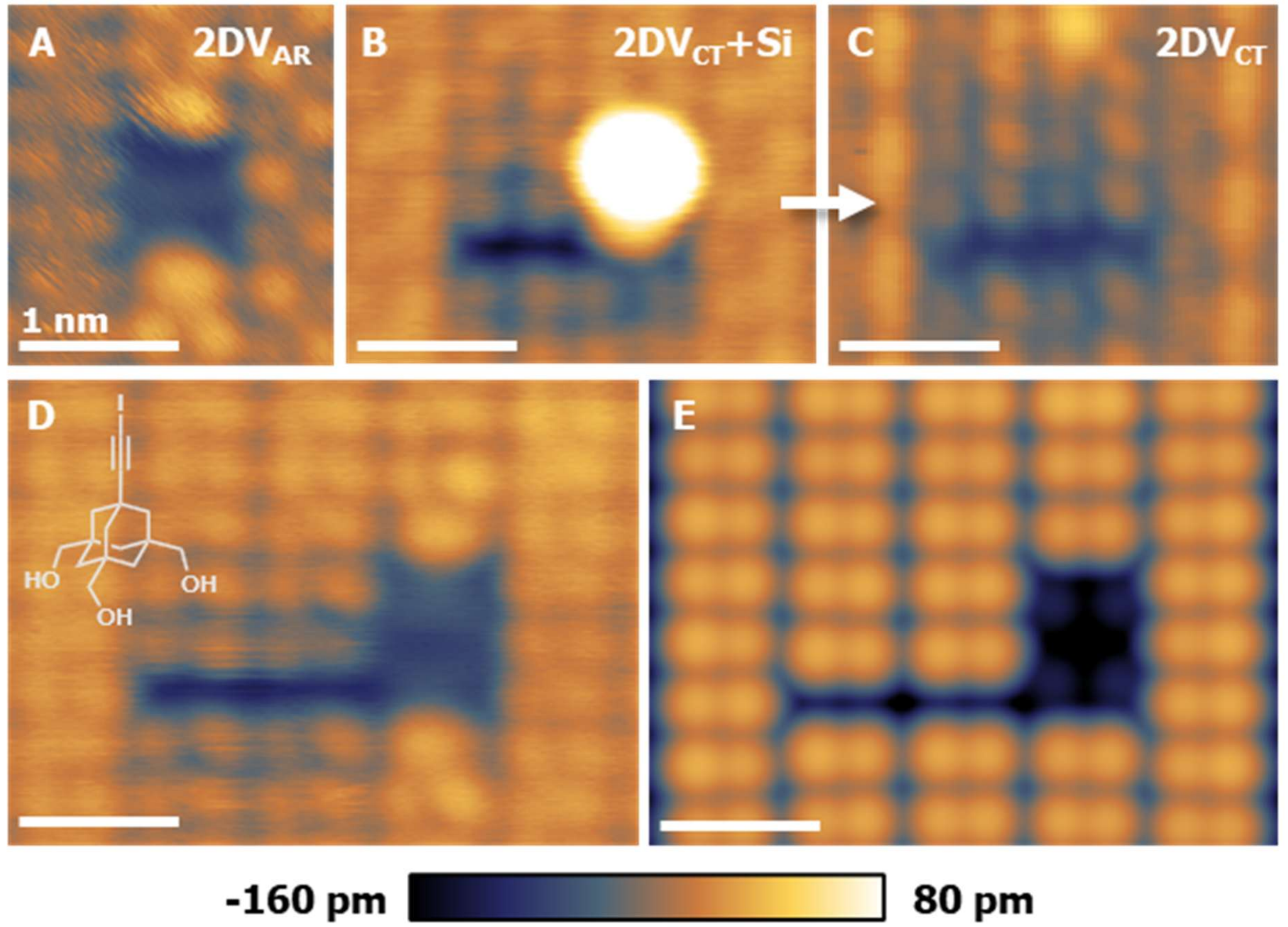

**Fig 5. Demonstration of subtractive silicon patterning at 77 K.** Along-row (**A**) and cross-trough (**B**) dimer vacancy pairs ($2DV_{AR}$ and $2DV_{CT}$). An example of extraneous silicon abstraction (**C**) and correcting the error to form the intended $2DV_{CT}$ feature. (**D**) A representative example of extended subtractive silicon patterning to form an "L" shape with inset representation of MAOC–$C_2I$, and (**E**) an STM simulation of a plausible final structure, simulated at −2.8 V sample bias with a 50 nA isocurrent. Images gathered in IM-STM under −3.4 V applied sample bias and 20 pA current tunneling conditions.

## Conclusions

Herein, we have demonstrated the process and utility of atomic mechanosynthesis *via* molecular tools as a method for APF in both additive and subtractive modes. The molecular tools described are of a general kind, typified by intentional synthesis for a certain range of behaviors and functionalities. The build site is highly flexible, with Si(100) presented here as a foundational build site of interest for its technological relevance and extensive existing studies in literature. Even the simplest *z*-only reaction trajectories with only pre-interaction lateral targeting were shown to produce previously-unreported surface arrangements of carbon and silicon. This technique of combining IM-STM and molecular tools exhibits clear paths towards more complex arrangements, additional transferable moieties, and generally enabling new APF methods as a prerequisite for eventual, atomically precise manufactured products.

The capabilities demonstrated here are not uniquely dependent on EAOGe–$C_2$I. One of its many possible variants, MAOC–$C_2$I, was demonstrated to be specifically suitable for subtractive patterning of silicon. The method was also shown to be robust at both 4 K and 77 K, with rich avenues for further study. Thus, nanotechnologies previously considered infeasible due to their dependence on each individual atom being placed in precisely the intended location with covalent bonding are brought closer to fruition by positionally-controlled mechanosynthesis as outlined herein. Future developments are expected to be made with respect to the use of mechanosynthetic building blocks beyond $C_2$ and more intricate trajectories to fully exploit the available parameter space for IM-STM enabled mechanosynthesis, among others.

**Acknowledgments:** We acknowledge the efforts of many other colleagues in their ongoing support of CBN Nano Technologies, Inc. as a whole, including Darian Blue, Byoung Choi and Sheldon Haird. We thank David Lister for discussions regarding IM-STM with molecular tools and early input on SPM head design. We also thank John Turriff, Mike Meakin, and Tom McKendree for project management inputs. No parts of the text or figures were drafted by AI-based tools, but Microsoft Copilot was utilized, based on human-written drafts, to simulate anticipated peer-review feedback reported by others (*30*). This feedback was then considered by the co-authors to further clarify the text as part of manuscript refinement.

**Funding:** This work was supported in part by the Canadian government under the Strategic Innovation Fund (SIF, Agreement Number 813022).

**Author contributions:**

Conceptualization: DGA, MD, AH, RAFJ, RM, M. Marshall, M. Moses, TT, RY

Methodology: DGA, EBR, AH, MD, TM, M. Morin, KT, MK

Investigation: AA, DGA, KSA, ZA, AB, BB, DC, KC, MC, RC, CD, TE, JZF, ATKG, RG, SYG, CI, RK, SL, AM, CM, HM, JM, M .Morin, M. Moses, OM, SM, TM, AP, RP, HR, SR, BS, LS, MS, DABT, BT, JLW, RW, CY, RY

Visualization: ZA, AB, BB, RC, RG, AI, CI, MJ, M.Morin, TM

Funding acquisition: JB, RAFJ, RM

Project administration: DGA, M. Drew, M. Durand, MK, MT, TT, DV, RY

Supervision: DGA, RA, BB, EBR, TH, AI, MK, M.Morin, M. Moses, OM, TT, KSA, MT, DV

Writing – original draft: DGA, ZA, AB, BB, RC, CI, MJ, CM, M. Morin, TM, MS, RW

Writing – review & editing: DGA, AB, BB, AH, CM, MS, CI, ZA, RC, MJ, M. Morin, TM, RW, RK, HM, OM, SR, MT, BS, JM

**Competing interests:** The authors declare competing financial interests; patents have been filed on aspects of this work. All authors are or were affiliated with CBN Nano Technologies, Inc. a company seeking to commercialize atomically precise 3D manufacturing. The work presented in this manuscript was also funded in part by the Canadian government under the Strategic Innovation Fund (SIF, Agreement Number 813022).

**Data, code, and materials availability:** The data underlying this work are not publicly available, as they constitute a part of CBN Nano Technology Inc.'s intellectual property. Reasonable, academic requests for data not included in this work or in the Supporting Information may be submitted to the corresponding author/s or info@nfcbn.com for consideration.